\pdfoutput=1

\documentclass[
  ,draft            
  ]
  {aipproc}

\layoutstyle{6x9}

\begin{document}

\title{Wrong Priors}


\author{Carlos C. Rodr{\'{\i}}guez} {
  address={ \url{http://omega.albany.edu:8008/} \\
  Department of Mathematics and Statistics \\
  The University at Albany, SUNY\\
  Albany, NY 12222\\
  USA}
}

\begin{abstract}
All priors are not created equal. There are right and there are wrong priors.
  That is the main conclusion of this contribution. I use, a cooked-up example
  designed to create drama, and a typical textbook example to show the pervasiveness
  of wrong priors in standard statistical practice. 
  
\end{abstract}

\maketitle


\section{Introduction}

  The information geometry available in regular statistical models can be used
to build objectively meaningful prior distributions. When the information volume
of the model is finite, the uniform distribution over the model manifold coincides
with Jeffreys invariant rule. A simple example in two dimensions shows that 
the popular naive diffuse prior over the parameters of this model is in fact a
\emph{wrong} prior, requiring more than ten thousand observations to match
Jeffreys rule with only 100 samples. Bayesian inference
is suffering from an epidemic of wrong priors and to prove it I consider 
standard simple logistic regression with naive diffuse priors and with uniform
priors over the manifold. The results are obviously less dramatic but 
similar to the previous cooked-up example. When the information volume of the model
is infinite, the uniform distribution over the model does not exist. However, the available
geometry can still be exploited and it provides a semiparametric family of invariant
objectively ignorant priors.

\section{A simple example}
\begin{figure}\label{fig:twins}
  \includegraphics[height=.5\textheight]{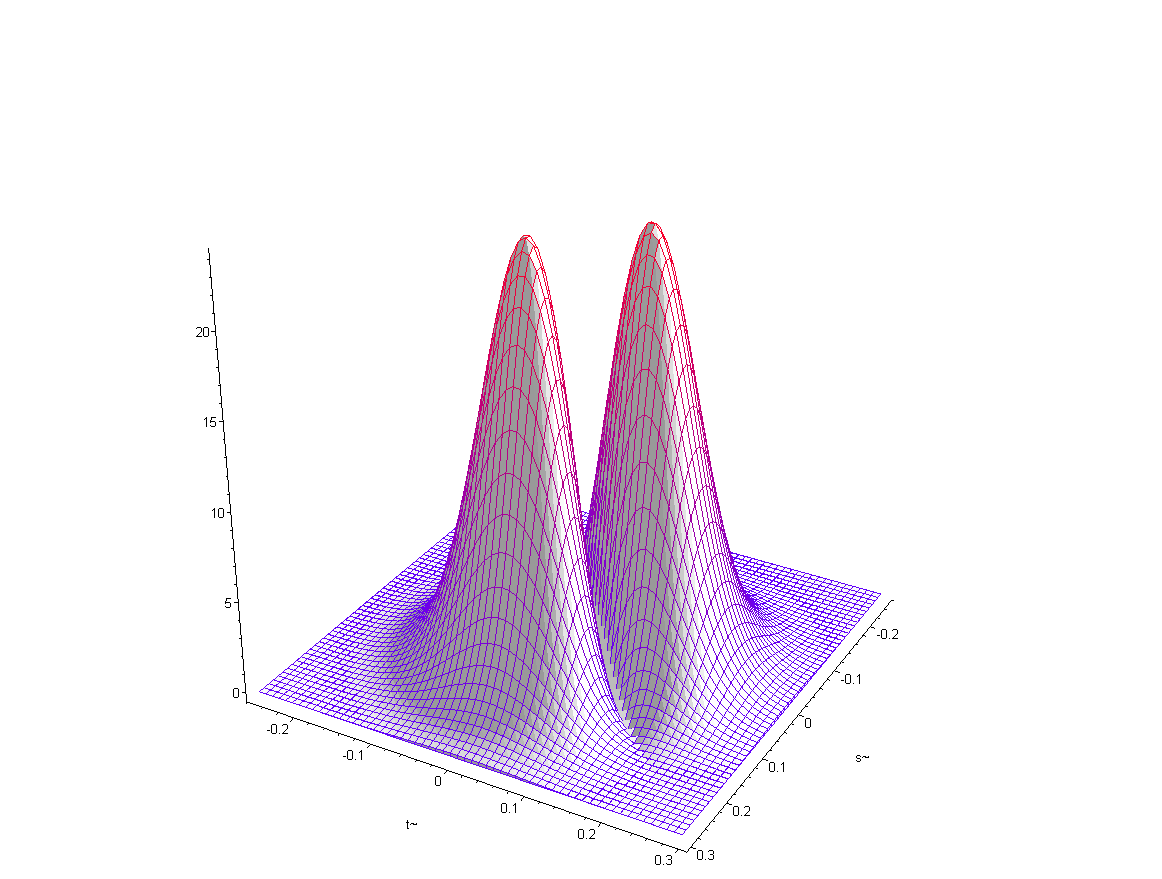}
  \caption{The prior $\pi_{1}$ is the true uniform over the model. Notice that the 
  picture is not drawn at scale. The actual peaks should be more than 40 times taller
  than the ones displayed. }
\end{figure}

Consider bivariate normals with unit covariance matrix
and mean vector restricted to a region of the euclidean plane. Specifically, for given values
$a$ and $b$ the experiment consists of choosing $(x,y)$ at random on the euclidean plane with,

\begin{eqnarray*}
	x &=& \exp\left(-50 ( a^{2} + b^{2} ) \right) + \epsilon_{1} \\
	y &=&  \exp\left( -50 ( (a-c)^{2} + (b-c)^{2} ) \right) + \epsilon_{2}. \\
\end{eqnarray*}
The unknown parameters are $a,b \in R$ but $c= 0.1$ is assumed known, and 
$\epsilon{_1}$ and $\epsilon_{2}$ are independent standard normals. The problem consists of learning the
parameters $\theta = (a,b)$ from $n$ independent observations $(x_{1},y_{1}),\ldots, (x_{n},y_{n})$. We want to compare the performance of two priors on $(a,b)$. The naive "`ignorant"' prior $\pi_{0}$ that takes $a$ and $b$ independently from $N(0,100)$ and the uniform prior over the manifold model, $\pi_{1}$ given by,

\begin{equation} \label{eq:pi1}
\pi_{1}(a,b) = \frac{|a-b|}{Z} \exp\left(-100(a^{2}+ b^{2}) + 10 (a+b) \right)
\end{equation}
where $Z$ is a finite normalization constant. Equation (\ref{eq:pi1}) is just the normalized volume form 
of the model computed trivially as $\sqrt{\det g}/Z$ with $g$ as the information matrix (minus the expected values of the second derivatives of the log likelihood).
This prior puts positive mass on the entire $(a,b)$ plane (except on the line $a=b$ but that region has measure 0) but it is very far from uniform as it is shown in figure \ref{fig:twins}. Notice also that there are two peaks because the likelihood is invariant under the exchange of $a$ with $b$. The volume prior respects this symmetry.


\subsection{Posterior Inference}
\begin{figure}\label{fig:mcmctwins}
  \includegraphics[height=0.6\textheight]{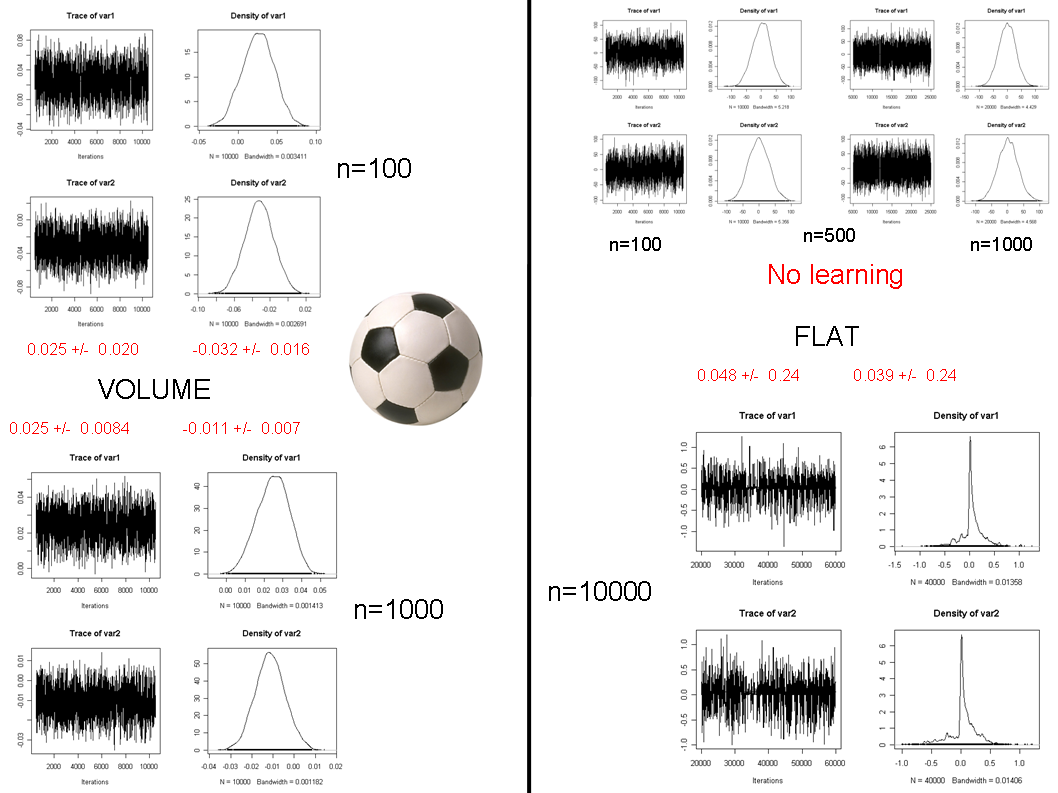}
  \caption{The right side shows the posteriors for the naive flat prior. The left side shows the posterior computed with the true uniform over the model. }
\end{figure}

With the help of the free MCMC package \cite{MCMCPack,RManual} it only takes a few lines of code to
realize the inadequacy of the naive prior for this example. The results of the MCMC simulations are summarized in figure \ref{fig:mcmctwins}.
The true parameters where fixed at $a = 0.025$ and $b = -0.01$ and independent samples were chosen from the distribution with those parameters. With the naive flat prior the posteriors after observing $100, 500$ and $1000$ samples were essentially identical to the priors $N(0,100)$, i.e. nothing was learned from the data. With $10000$ observations the program was
able to learn the values $(0.048 \pm 0.24, 0.039 \pm 0.24)$ for the true parameters. In contrast, just after
$100$ observations the posterior with the true uniform prior estimates the parameters very precisely as 
$(0.025 \pm 0.020, -0.032 \pm 0.016)$, still one order of magnitude of extra accuracy over the posterior with
the flat prior with two orders of magnitude of extra data!

\subsubsection{Why is the volume prior so good?}
\begin{figure}\label{fig:leafradius3}
  \includegraphics[height=.4\textheight]{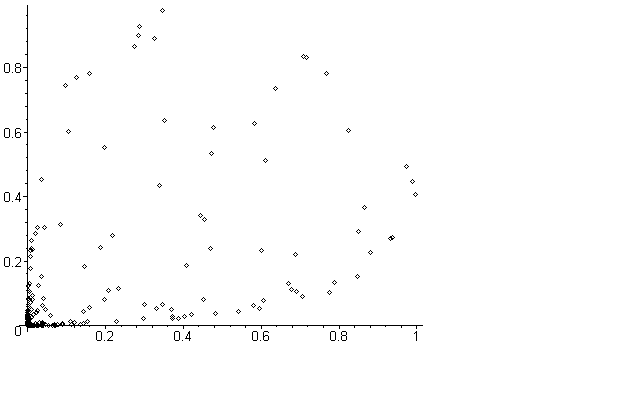}
  \caption{Ten thousand $(u,v)$ points from $(a,b)$ points chosen uniformly inside a ball of radius 3 centered at $0$. Notice that more than $9000$ points disappear into the origin.}
\end{figure}

\begin{figure}\label{fig:priorsamples}
  \includegraphics[height=.4\textheight]{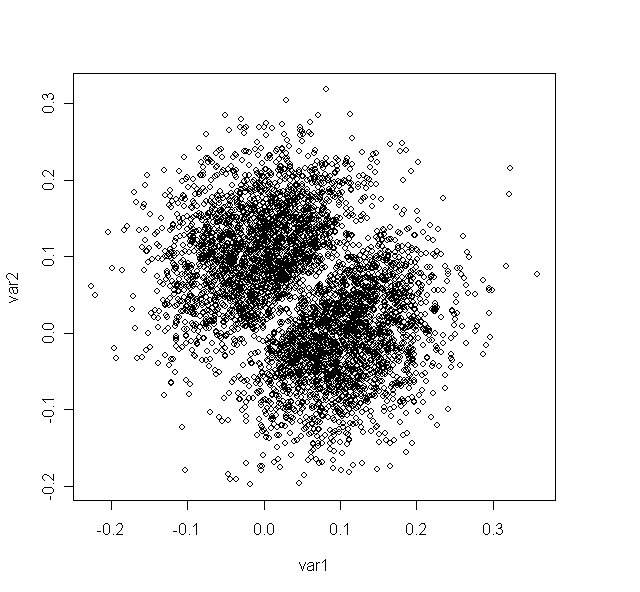}
  \caption{Ten thousand $(a,b)$ points from the true uniform over the model}
\end{figure}

\begin{figure}\label{fig:unifleaf}
  \includegraphics[height=.4\textheight]{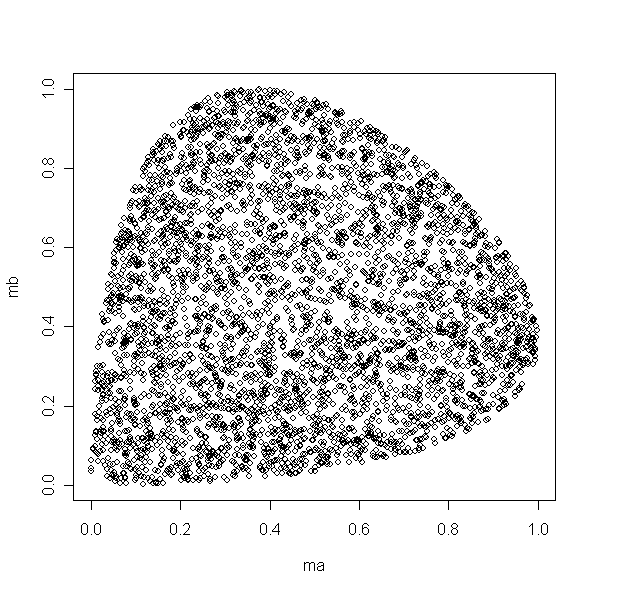}
  \caption{Ten thousand $(u,v)$ points from $(a,b)$ chosen from the density (\ref{eq:pi1}).}
\end{figure}

To understand why the naive flat prior is so bad and the volume prior so good let's identify the transformed region of means $(u,v)$ given by,

\begin{eqnarray*}
	u &=& \exp\left(-50 ( a^{2} + b^{2} ) \right) \\
	v &=&  \exp\left( -50 ( (a-c)^{2} + (b-c)^{2} ) \right) \\
\end{eqnarray*}
as $(a,b)$ range over the entire plane. An easy way to find the shape of this region is to pick points $(a,b)$ at random on the plane and plot the corresponding $(u,v)$ points. Figure \ref{fig:leafradius3} shows $10000 (u,v)$  points obtained from $10000 (a,b)$ points uniformly distributed inside a circle centered at the origin of radius $3$. Notice that lots of points disappear into the origin!. Now take another $10000 (a,b)$ points but now distributed according to $\pi_{1}$ with density given in (\ref{eq:pi1}). 
Figure \ref{fig:priorsamples} shows these $(a,b)$ points. Notice that they are all highly concentrated about two points close to the origin. The corresponding $(u,v)$ points are shown in figure \ref{fig:unifleaf}. Got it?

\subsubsection{The equation of the boundary of the leaf of $(u,v)$ points}

\begin{figure}\label{fig:Theleaf}
  \includegraphics[height=.3\textheight]{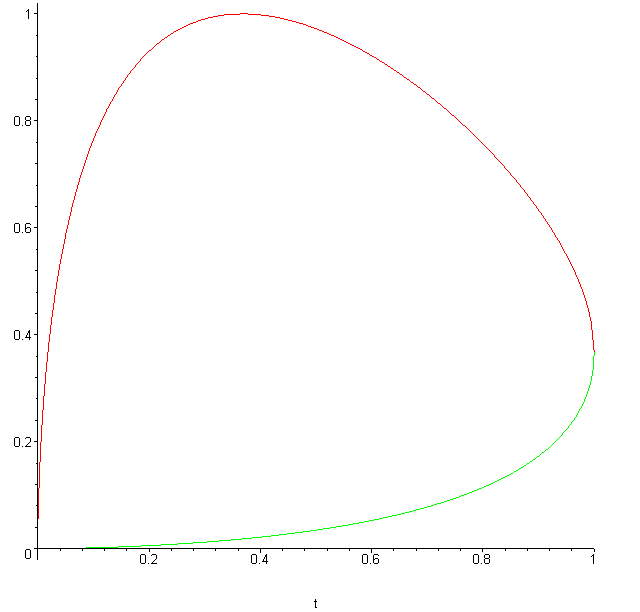}
  \caption{The exact boundary of the $(u,v)$ points. Upper dark (red) part is $R(t)$. Lower light (green) part is $G(t)$}
\end{figure}

The computation of the exact equation of the leaf boundary in figure \ref{fig:unifleaf} is a nice
exercise in simple optimization: Find max and min of $v$ subject to the
constraint that $u=t$. The max is given by the Red (dark) $R(t)$ curve in figure \ref{fig:Theleaf}, with

\begin{equation}\label{eq:Rt}
R(t) = \exp(- (\sqrt{-\log t}-1)^{2} ).
\end{equation}

The min is given by the Green (light) curve,

\begin{equation}\label{eq:Gt}
 G(t) = \exp( - (\sqrt{-\log t}+1)^{2} )
\end{equation}
with $0<t<1$ in both cases.

Notice that there is a non-removable corner singularity at $t=0$ but it
is a piece of euclidean space so the curvature is zero at every
point.

\begin{table}
	\centering
		\begin{tabular}{ccc}
		\hline
		Dose, $x_{i}$ &   Number of   &   Number of \\
		($\log$\ g/ml) &  animals, $n_{i}$ & deaths, $y_{i}$ \\
		  \hline
		-0.863    &    5   &   0 \\
		-0.296    &    5   &   1 \\
		-0.053    &    5   &   3 \\
		 0.727    &    5   &   5 \\
		  \hline
		\end{tabular}
	\caption{Bioassay data from  \cite[p.88]{BDA04}}
	\label{tab:gelman}
\end{table}

\section{Textbook example}

	Perhaps the first non-trivial example of a multiparameter bayesian model is simple logistic regression (see \cite[p.88]{BDA04}). Twenty animals were tested, five at each of four dose levels (see Table \ref{tab:gelman}). The standard model for this kind of data is,

\[ (x_{i},n_{i},y_{i}); \ i = 1,\ldots,k, \]
assumed independent with,
\[
  y_{i} | \theta_{i} \sim \mbox{Bin}(n_{i},\theta_{i}),
  \]
where $\theta_{i}$ is the probability of death for animals given dose $x_{i}$. The standard logistic dose-response relation is:

\begin{equation}\label{eq:logreg}
 \log \frac{\theta_{i}}{1-\theta_{i}} = a + b x_{i}
\end{equation}

\begin{figure}\label{fig:p4logist}
  \includegraphics[height=.5\textheight]{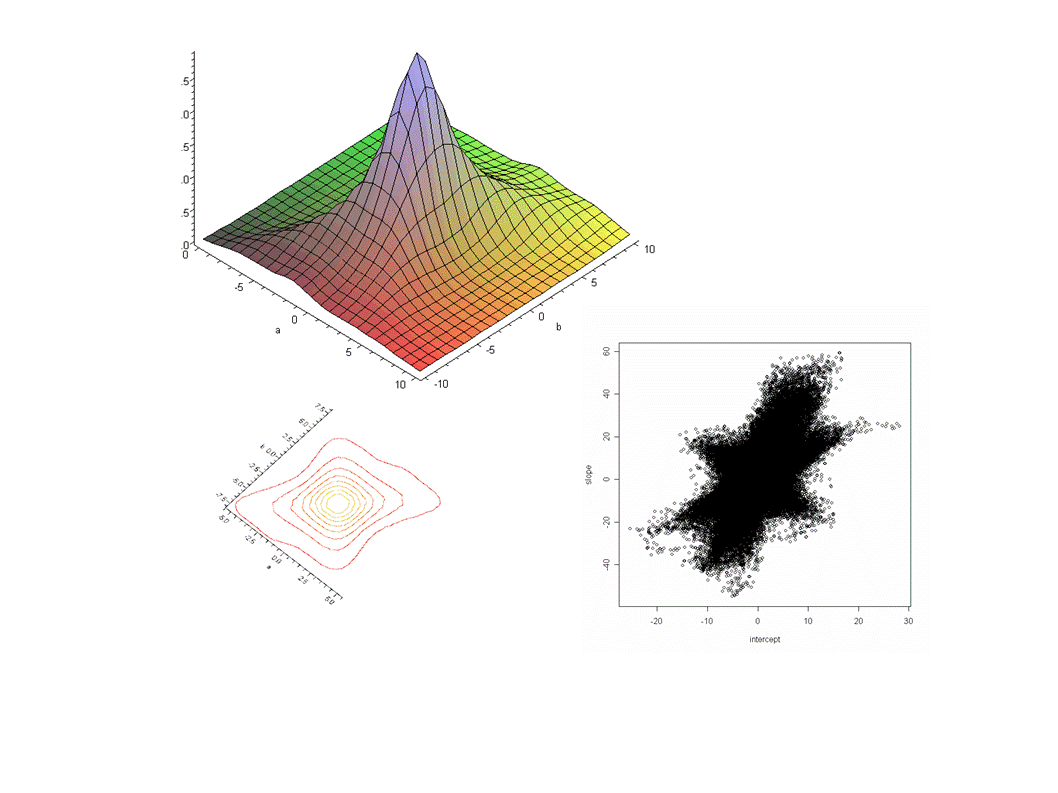}
  \caption{Uniform prior for logistic regression in the $(a,b)$ parameterization. Bottom left: Contours. Bottom right: $2.5 \times 10^{5}$ samples from this prior.}
\end{figure}

\begin{figure}\label{fig:pos4logist}
  \includegraphics[height=.5\textheight]{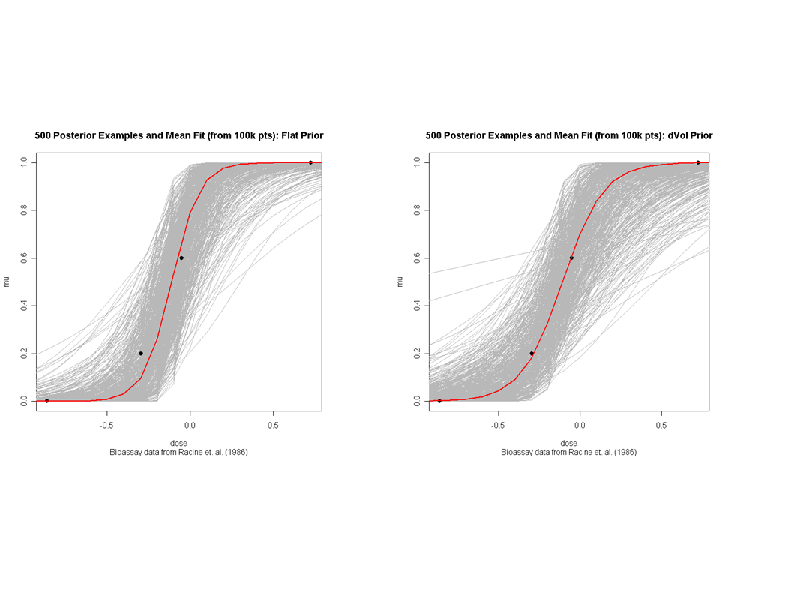}
  \caption{Posterior logistic curves. Left panel with naive prior. Right panel with volume prior.}
\end{figure}

The joint distribution of $(y_{1},\ldots,y_{k})$ is a function of the unknown parameters $(a,b)$ and straight (but tedious) calculations give the volume element $dV = \sqrt{\det g}\ da db$ in the $(a,b)$ parameterization as

\begin{eqnarray*}
dV &=& T\ \sigma\ da db \\
T  &=& \sum_{j} w_{j} = \sum_{j} n_{j}\theta_{j}(1-\theta_{j}) \\
\sigma &=& \mbox{ stand. dev. of\ } X  \mbox{\ defined as,} \\
  &\ &   P\left\{ X = x_{j} \left| \theta_{j} \right. \right\} \propto w_{j} \\
\theta_{j} &=& \frac{1}{1+ \exp(-a-b x_{j})} \\
\end{eqnarray*}
This is a strange looking density (see figure \ref{fig:p4logist}). In particular this prior is proper and it assigns correlation of about $0.5$ between $a$ and $b$. This correlation is known a priori from the underlying geometry. In fact, the volume prior provides a better fit to the data than the standard diffuse naive prior that models $a$ and $b$ as independent variables with large variances. Figure \ref{fig:pos4logist} shows the results of the posterior simulations with both priors. Left panel with naive prior, right panel with volume prior. The red (dark) middle curves represent the logistic curves associated to the mean posterior values for $(a,b)$ (100 thousand of them). The pictures also show 500 logistic curves obtained by sampling 500 $(a,b)$ pairs from the available posterior samples. There is clearly more spread of logistic curves on the right than on the left panel. This is compatible with the fact that the volume prior samples uniformly over the manifold. Just like in the cooked-up example the over-spread $(a,b)$ points cover only a small region of the manifold.

\section{Beyond finite volumes}
When the information volume $V(M)$ of the model $M$,

\[ V(M) = \int_{M} dV = \int_{\Theta} \sqrt{\det g}\ d\theta \]
is infinite; there is no uniform distribution over $M$. However, the underlying information geometry provides the following class of priors given as scalar density fields defined invariantly on $M$ by,

\begin{equation}\label{eq:gprior}
\pi(p |t,\nu,\delta,\alpha) = \frac{1}{Z} [ 1 + \alpha \nu I_{\delta}(p:t) ]^{- \frac{1}{\nu}}
\end{equation}
where $p\in M$, $t$ is a probability distribution guessing the actual distribution of the data, $\delta,\nu$ are scalar parameters in $[0,1]$, $\alpha > 0$ large enough so that $Z < \infty$ and $I_{\delta}(p:t)$ is the $\delta$-information deviation between (unnormalized) distributions $p$ and $t$ given by,

\[ I_{\delta}(p:t) = \frac{1}{\delta (1-\delta)} \int [\delta p + (1-\delta) t - p^{\delta} t^{1-\delta} ] \]
where the integral is over the whole data space manifold. This family of priors exists for any regular model and it has many remarkable properties. In particular this family maximizes a simple and objective notion of ignorance. For details see my \emph{ A geometric theory of ignorance }. The hyper parameters can be estimated with priors of the same kind or with a nonparametric prior of the Dirichlet Process type (which could itself be seen as part of this family if we allow $M$ to be infinite dimensional). There are still many open problems but the road ahead seems clear: More geometry.


\begin{theacknowledgments}
  I am in debt to Phil Dawid whose invitation to talk at UCL prompted the finding of the example in the paper and to Ariel Caticha, Kevin Knuth, and John Skilling for many interesting discusions.
\end{theacknowledgments}



\bibliographystyle{aipproc}   

\bibliography{carlos}

\IfFileExists{\jobname.bbl}{}
 {\typeout{}
  \typeout{******************************************}
  \typeout{** Please run "bibtex \jobname" to optain}
  \typeout{** the bibliography and then re-run LaTeX}
  \typeout{** twice to fix the references!}
  \typeout{******************************************}
  \typeout{}
 }

\end{document}